\documentclass[a4paper]{jpconf}
\usepackage{graphicx}
\begin{document}




\title{Crystal growth and magnetic properties of equiatomic CeAl}
\author{Pranab Kumar Das\footnote[1]{Present address: Istituto Officina dei Materiali (IOM) - CNR, Laboratorio TASC, Area Science Park, S.S.14, Km 163.5, I-34149 Trieste, Italy. } and A. Thamizhavel}
\address{Department of Condensed Matter Physics and Materials Science, \\Tata Institute of Fundamental Research, Colaba, Mumbai 400 005, India}
\ead{thamizh@tifr.res.in}

\begin{abstract}

Single crystal of CeAl has been grown by flux method using Ce-Al self-flux. Several needle like single crystals were obtained and the length of the needle corresponds to the [001] crystallographic direction. Powder x-ray diffraction revealed that CeAl crystallizes in orthorhombic CrB-type structure with space group $Cmcm$ (no. 63). The magnetic properties have been investigated by means of magnetic susceptibility, isothermal magnetization, electrical transport, and heat capacity measurements. CeAl is found to order antiferromagnetically with a N$\grave{\rm e}$el temperature $T_{\rm N}$ = 10~K. The magnetization data below the ordering temperature reveals two metamagentic transitions for fields less than 20~kOe. From the inverse magnetic susceptibility an effective moment of $2.66~\mu_{\rm B}$/Ce has been estimated, which indicates that Ce is in its trivalent state. Electrical resistivity data clearly shows a  sharp drop at 10~K due to the reduction of spin disorder scattering of conduction electrons thus confirming the magnetic ordering. The estimated residual resistivity ratio (RRR) is 33, thus indicating a good quality of the single crystal. The bulk nature of the magnetic ordering is also confirmed by heat capacity data.  From the Schottky anomaly of the heat capacity we have estimated the crystal field level splitting energies of the $(2J+1)$ degenerate ground state as 25~K and 175~K respectively for the fist and second excited states.
\end{abstract}

\section{Introduction}
Magnetism in rare earth intermetallic compounds is interesting due to the competing interaction between Ruderman-Kittel-Kasuya-Yosida (RKKY) and Kondo exchange interaction. Such compounds exhibit various diverse ground states depending on the relative strength of RKKY and Kondo exchange interaction, for example, CeAl$_3$, CeCu$_6$~\cite{Andres, Onuki} are heavy-fermion Kondo systems without long range magnetic ordering; CeAl$_2$, CeB$_6$~\cite{Barbara, Winzer} are Kondo lattice compounds with magnetic ordering at low temperature.  In CeIn$_3$, pressure induced superconductivity is observed~\cite{Grosche}. In view of these interesting properties, rare earth intermetallic compounds, especially Ce-based compounds, where $4f$ level lies close to the Fermi energy, have been studied extensively over last several decades. We have previously grown the single crystal of CeMg$_3$, and studied its magnetic properties~\cite{Pranab}. It was found that CeMg$_3$ is a heavy fermion Kondo lattice compound with a N\'{e}el temperature of 2.6~K. In continuation to our investigations on binary Ce compounds, we report here on the magnetic properties of the equiatomic compound CeAl. From the neutron scattering studies on polycrystalline CeAl Lawrence et al.,~\cite{Lawrence} have reported that this compound orders antiferromagnetically at 10.2~K, without any suppression of the local moment in the ground state.  They have estimated an ordered moment of 1.72~$\mu_{\rm B}$/Ce, which is very close to the free ion value of Ce$^{3+}$ (2.14~$\mu_{\rm B}$/Ce).  

\section{Experiment}

\begin{figure}[!]
\centering{}
\includegraphics[width=0.9\textwidth]{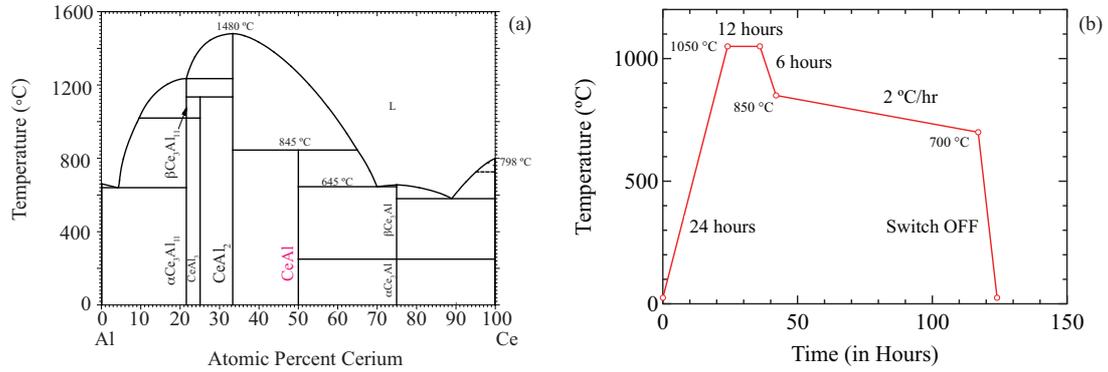}
\caption{\label{Fig1}(Colour online) (a) Binary phase diagram of Ce and Al extracted from Ref.~\cite{Van}.  (b)  Temperature profile adopted for the crystal growth of CeAl single crystal.}
\end{figure}
The binary phase diagram of Ce and Al is shown in Fig.~\ref{Fig1} which is extracted from Ref.~\cite{Van}.  It is obvious from the figure that the phase CeAl melts incongruently and hence could not be grown directly from its stoichiometric melt.  Hence the single crystals of CeAl have been grown from an off-stoichiometric ratio by the high temperature solution growth or the so called flux method.  High purity starting elements of $3N$ pure-Ce and $4N$ pure-Al were taken in the ratio $2:1$ in a high quality recrystallized alumina crucible and sealed inside a quartz ampoule under a vacuum of 10$^{-6}$~Torr.  The ampoule was heated to 1050~$^{\circ}$C in a resistive heating box-type furnace and held at this temperature for 24~hrs for proper homogenization of the melt.  The temperature profile of the crystal growth process is shown in Fig.~\ref{Fig1}(b).  In order to avoid the CeAl$_2$ phase formation, we cooled the melt very rapidly to 850~$^{\circ}$C and then employed a cooling rate of 2~$^{\circ}$C/hr down to 700~$^{\circ}$C at which point the  excess flux was centrifuged.  We have adopted similar procedure for the growth of LaAl single crystals as well. Several needle like crystals were obtained with typical dimensions of length about 1 to 2~mm and 0.5~mm thickness. CeAl single crystals were found to be stable in air while LaAl samples were found to be oxidizing when exposed in air for an extended period of time. The phase purity of the sample was determined from the powder x-ray diffraction measurement using PANalytical X-ray diffractometer with the monochromatic Cu$K_{\rm \alpha}$ radiation ( = 1.5406~${\rm \AA}$).  The electrical resistivity was measured down to 1.8~K by four probe method in a home made set-up. Magnetization and magnetic susceptibility measurements were performed using a Quantum Design superconducting quantum interference device (SQUID) magnetometer. Heat capacity was measured using a Quantum Design Physical Property Measurement System (PPMS).  

\section{Resutls and discussion}
\subsection{X-ray studies}
\begin{figure}[!]
\centering{}
\includegraphics[width=0.55\textwidth]{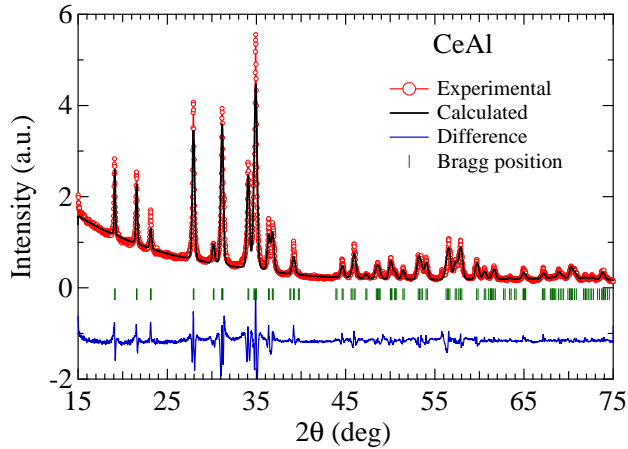}
\caption{\label{Fig2}(Colour online) Powder x-ray diffraction pattern of CeAl together with the Rietveld analysis.}
\end{figure}

Since the single crystals of CeAl have been grown from an off-stoichiometric starting composition, in order to check the phase purity, we recorded the x-ray diffraction (XRD) pattern of the powdered single crystal which is shown in Fig.~\ref{Fig2}. It is evident from the XRD pattern that the grown crystals are of single phase and no traces of any flux impurity is observed.  To estimate the lattice constant we performed the Rietveld refinement using FullProf software package~\cite{Fullprof}. The experimental and calculated patterns are shown in Fig.~\ref{Fig2}. The estimated lattice constants are: $a = 9.2903(4)~{\rm \AA}$, $b = 7.6840(3)~{\rm \AA}$, and $c = 5.7365(3)~{\rm \AA}$. The lattice parameters match well with the previously reported values~\cite{Asmat}. Here Ce occupies $8g$ Wyckoff's position with the positional parameters $x = 0.321(1)$, $y = 0.161(1)$; while Al possesses two different sites $4a$ and $4c$ with $y = 0.318(2)$. Energy Dispersive X-ray analysis (EDX) on a cross-sectional surface of needle-like crystal was performed and the stoichiometry of Ce and Al is found to be 1 : 1, which is the desired composition.

\subsection{Magnetization Studies}

\begin{figure}[h]
\centering{}
\includegraphics[width=0.75\textwidth]{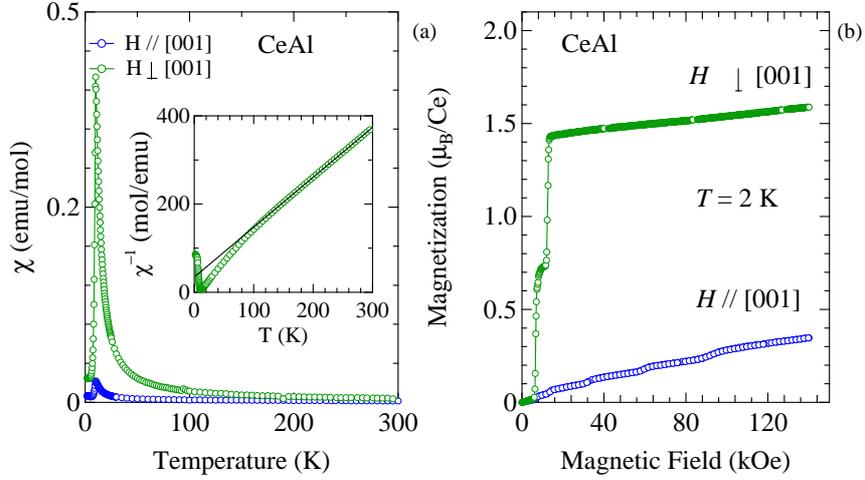}
\caption{\label{Fig3}(Colour online) (a) Temperature dependence of magnetic susceptibility measured parallel and perpendicular to the length of the crystal.  The inset of (a) depicts the inverse susceptibility plot.  (b) The isothermal magnetization of CeAl measured at 2~K along the two different crystallographic directions as mentioned in the figure.}
\end{figure}

Figure~\ref{Fig3} shows the temperature dependence of magnetic susceptibility of CeAl in the temperature range 1.8 to 300~K measured in an applied magnetic field of 1~kOe.  Since the crystals were very tiny we measured the magnetic susceptibility parallel and perpendicular to the length of the crystal.  From Laue diffraction, we found that the length of the crystal corresponds to [001] direction.   At high temperature the magnetic susceptibility follows Curie-Weiss behaviour.  A very sharp transition is observed at 10~K signalling the magnetic ordering in this compound.  This type of huge drop in the magnetic susceptibility indicates that the magnetic ordering is antiferromagnetic in nature.  In a two sub-lattice antiferromagnet along the easy axis of magnetization the susceptibility falls rapidly to zero while it remains almost constant along 
the perpendicular direction, namely the hard axis.  The inset of Fig.~\ref{Fig3}(a) shows the 
representative inverse susceptibility along the [001] direction.  The solid line is a fit to the Curie-Weiss law.  From the fitting we have estimated the effective magnetic moment $\mu_{\rm eff}$ and the paramagnetic Weiss temperature $\theta_{\rm p}$ as 2.66~$\mu_{\rm B}$/Ce and $-30$~K, respectively.  The obtained effective magnetic moment value is close to the free-ion value, indicating that Ce atom is in trivalent state.  The negative value of $\theta_{\rm p}$ is in conformity with the antiferromagnetic nature of magnetic ordering in CeAl.

Figure~\ref{Fig3}(b) shows the isothermal magnetization measured at 2~K.  Two metamagnetic transitions are observed due to the spin reorientation caused by the applied magnetic field.  It is evident from the figure that for fields parallel to the length of the crystal, the magnetization is linear and much smaller and attains only a value of 0.34~$\mu_{\rm B}$/Ce.  On the other hand, when the magnetic field is applied perpendicular to the length of the crystal, the magnetization undergoes two spin-flop transitions due to the re-orientation of the spins.  The first metamagnetic transition occurs at a critical field of 6~kOe followed by another sharp metamagnetic transition at 11~kOe.  The magnetization saturates at 13~kOe and attains a value of 1.6~$\mu_{\rm B}$/Ce at 140~kOe which is close to the expected to the saturation moment of Ce$^{3+}$  $(g_J J(=\frac{6}{7}\times\frac{5}{2})$ 2.14~$\mu_{\rm B}$/Ce.  The observed saturation moment is in conformity with the  previous neutron diffraction experiment on a polycrystalline sample~\cite{Lawrence}. 

\subsection{Resistivity Measurement}
\begin{figure}[b]
\centering{}
\includegraphics[width=0.5\textwidth]{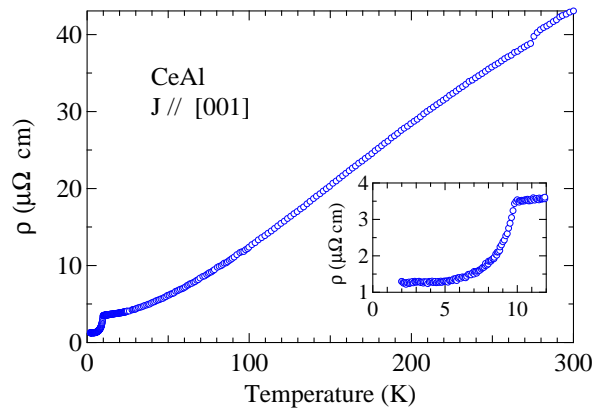}
\caption{\label{Fig4}(Colour online) Temperature dependence of electrical resistivity of CeAl measured in the temperature range from 1.8 to 300~K.  The inset shows the low temperature part of the electrical resistivity.}
\end{figure}

The electrical resistivity of CeAl measured along the length of the crystal, in the temperature range from 1.8 to 300~K is shown in Fig.~\ref{Fig4}, the inset shows the low temperature part of the electrical resistivity.  The resistivity decreases as the temperature is decreased and it shows a typical metallic behaviour.   At 10~K the electrical resistivity shows a sudden change of slope and decreases more rapidly due to the reduction in the spin disorder scattering caused by the antiferromagnetic ordering of the Ce moments.  The electrical resistivity attains a value close to 1~$\mu \Omega$~cm at the lowest temperature measured.  The residual resistivity ratio (RRR) is estimated to be 37 which indicates the high quality of the single crystal.  

\subsection{Heat Capacity Studies}

\begin{figure}[b]
\centering{}
\includegraphics[width=1\textwidth]{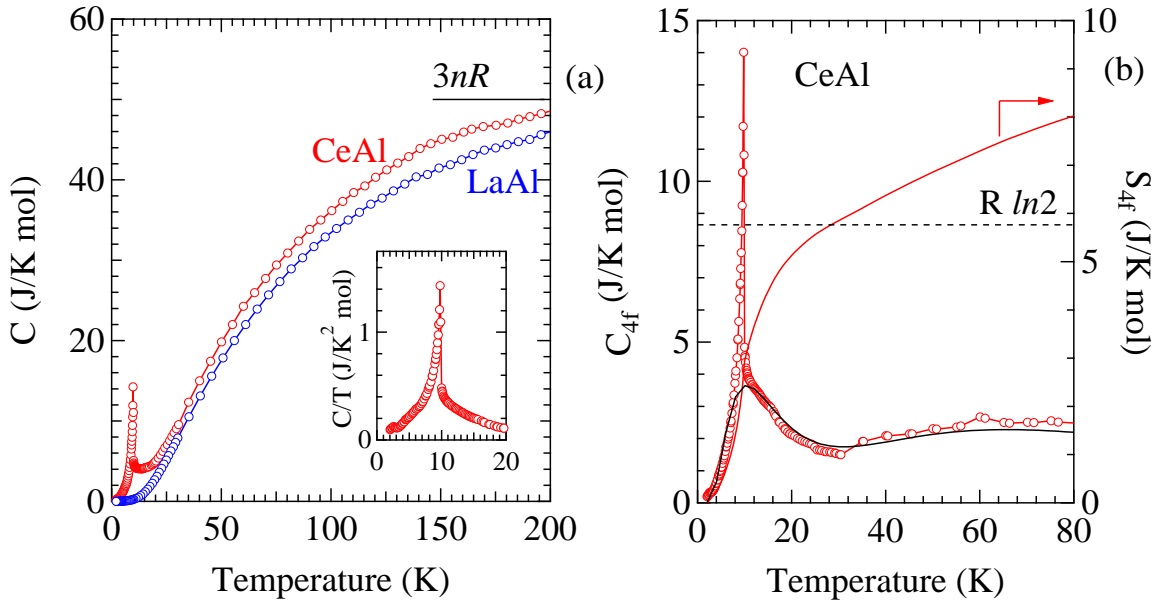}
\caption{\label{Fig5}(Colour online) (a) Temperature dependence of the specific heat of CeAl and that of LaAl single crystals.  The inset shows the low temperature part of the $C/T$ $vs.$ $T$ plot.  (b) The magnetic part of the heat capacity and the estimated magnetic entropy.  The black solid line shows the Schottky heat capacity (see text for details).}
\end{figure}

The specific heat capacity of CeAl and that of the non-magnetic reference compound LaAl measured in the temperature range from 1.8 to 200~K is shown in Fig.~\ref{Fig5}(a).  A very sharp peak at 10~K suggests the bulk nature of magnetic ordering.  The specific heat of LaAl is typical of a metallic sample. From the low temperature part of the heat capacity data, we have estimated the Sommerfeld constant $\gamma$ as 52~mJ/K~mol for CeAl and 2~mJ/K~mol for LaAl. The inset shows the low temperature part of $C/T~vs~T$ plot.  The magnetic part of the heat capacity of CeAl is obtained by subtracting the heat capacity of LaAl and is shown in Fig.~\ref{Fig5}(b).  The jump in the $C_{\rm 4f}$ amounts to 12.48~J/K~mol as expected for a spin-half system in the mean field model.  This indicates that the Kondo effect is not significant in this compound and the magnetic ordering is due to the Ruderman-Kittel-Kasuya-Yosida (RKKY) interaction mediated by the conduction electrons.   The magnetic entropy $S_{\rm 4f}$ obtained by integrating the $C_{\rm 4f}/T$ is also shown in Fig.~\ref{Fig5}(b).  The entropy reaches $R~ln2$ just above the magnetic ordering indicating that the ground state is a doublet.  The magnetic part of the heat capacity shows a Schottky type anomaly above $T_{\rm N}$.  We have estimated the Schottky heat capacity and found that the $2J+1$ degenerate levels of Ce split into three doublets. The black solid line in Fig~\ref{Fig5}(b) shows the calculated Schottky heat capacity. The splitting energies obtained are 0, 25 and 175~K corresponding to the ground, first and second excited states.    At 200~K, the experimental heat capacity reaches a value of about 48~J/K~mol, which is close to the Dulong-Petit limiting value of $3nR~=~49.884$~J/K~mol.  

\section{Conclusion}

We have grown the single crystals of CeAl and LaAl by self flux method.   Needle like crystals with well defined morphology were obtained.  From the magnetic and transport measurements we found that this compound undergoes an antiferromagnetic transition at 10~K.  For $H~\parallel$~[001] direction the magnetization is linear while along the perpendicular direction a two step metamagnetic transition is observed indicating the easy axis of magnetization.  The electrical resistivity also showed a sharp drop  at 10~K conforming the magnetic ordering and the residual resistivity is close to 1~$\mu \Omega$~cm suggests a good quality of the single crystal.  The jump in the magnetic part of the heat capacity was estimated to be 12.45~J/K~mol which is typically observed for a spin-half system in the mean field model.  A Schottky type anomaly was observed in $C_{\rm 4f}$, immediately above the magnetic ordering and we have estimated the crystal field level splitting as 0, 25 and 175~K.  

\section{Acknowledgement}

We thank Mrs. Ruta Kulkarni for her help in measuring the electrical resistivity.

\end{document}